\documentclass[aps,prd,twocolumn,showpacs,amsmath,amssymb]{revtex4-1}
\usepackage{amsmath}
\usepackage{graphicx}
\usepackage{subfigure}
\usepackage{epstopdf}
\usepackage{color}
\usepackage{multirow}
\usepackage{setspace}
\usepackage{overpic}
\usepackage{amssymb}
\usepackage[bookmarksnumbered, pdfstartview=FitH,colorlinks,urlcolor=blue, citecolor=blue,linkcolor=blue] {hyperref}
\usepackage{lineno}
\usepackage{bm}
\usepackage{rotating}
\usepackage{ulem}
\usepackage[utf8]{inputenc}

\makeatletter
\renewcommand{\@thesubfigure}{\hskip\subfiglabelskip}
\makeatother

\let\oldequation\equation
\let\oldendequation\endequation

\renewenvironment{equation}
{\linenomathNonumbers\oldequation}
{\oldendequation\endlinenomath}

\begin{document}
	
	\author{
		\begin{small}
			\begin{center}
				M.~Ablikim$^{1}$, M.~N.~Achasov$^{11,b}$, P.~Adlarson$^{70}$, M.~Albrecht$^{4}$, R.~Aliberti$^{31}$, A.~Amoroso$^{69A,69C}$, M.~R.~An$^{35}$, Q.~An$^{66,53}$, X.~H.~Bai$^{61}$, Y.~Bai$^{52}$
% ...
				\\
				\vspace{0.1cm}
				(BESIII Collaboration)\\
				\vspace{0.1cm}
				{\it
					$^{1}$ Institute of High Energy Physics, Beijing 100049, People's Republic of China\\
% ...
$^{4}$ Bochum Ruhr-University, D-44780 Bochum, Germany\\
% ...
$^{11}$ G.I. Budker Institute of Nuclear Physics SB RAS (BINP), Novosibirsk 630090, Russia\\
% ...
$^{31}$ Johannes Gutenberg University of Mainz, Johann-Joachim-Becher-Weg 45, D-55099 Mainz, Germany\\
% ...
$^{35}$ Liaoning Normal University, Dalian 116029, People's Republic of China\\
% ...
$^{52}$ Southeast University, Nanjing 211100, People's Republic of China\\
					$^{53}$ State Key Laboratory of Particle Detection and Electronics, Beijing 100049, Hefei 230026, People's Republic of China\\
% ...
$^{61}$ University of Jinan, Jinan 250022, People's Republic of China\\
% ...
$^{66}$ University of Science and Technology of China, Hefei 230026, People's Republic of China\\
% ...
$^{69}$ University of Turin and INFN, (A)University of Turin, I-10125, Turin, Italy; (B)University of Eastern Piedmont, I-15121, Alessandria, Italy; (C)INFN, I-10125, Turin, Italy\\
					$^{70}$ Uppsala University, Box 516, SE-75120 Uppsala, Sweden\\
% ...
$^{b}$ Also at the Novosibirsk State University, Novosibirsk, 630090, Russia\\
% ...
				}
			\end{center}
			\vspace{0.2cm}
		\end{small}
	}

	\title{\boldmath Measurement of branching fraction of $D^{*+}_s\to D^+_s \pi^0$ relative to $D^{*+}_s\to D^+_s  \gamma$}
	
	\begin{abstract}
		
		Based on 7.33~fb$^{-1}$ of $e^+e^-$ collision data taken at center-of-mass energies between 4.128 and 4.226~GeV with the BESIII detector, we measure
		the branching fraction of $D^{*+}_s\to D^+_s\pi^0$ relative to that of $D^{*+}_s\to D^+_s\gamma$ to be
		$(6.16\pm 0.43\pm 0.19)\%$.
		The first uncertainty is statistical and the second one is systematic.
		By using the world average value of the branching fraction of $D^{*+}_s\to D^+_se^+e^-$,
		we determine the branching fractions of $D^{*+}_s\to D^+_s\gamma$ and $D^{*+}_s\to D^+_s\pi^0$
		to be $(93.57\pm0.44\pm0.19)\%$ and $(5.76\pm0.44\pm0.19)\%$, respectively.
	\end{abstract}
	\maketitle

	\section{Introduction}
	
	The excited strange charmed meson, $D_s^{*+}$, is formed from $c\bar{s}$ quark-antiquark pair.
	Throughout this paper, charge-conjugate states are always included.
	The $D_s^{*+}$ decays are dominated by the radiative process $D_s^{*+}\to D_s^+\gamma$ and the isospin-violating hadronic process $D_s^{*+}\to D_s^+\pi^0$ due to the quark SU(2) flavor breaking and isospin violating effects.
	Measurements of the branching fractions (BFs) of the $D_s^{*+}$  decays are important to explore
	quantum chromodynamics~(QCD)~\cite{Fritzsch:1973pi} describing the strong interaction.
	The decay widths of $D_s^{*+}\to D^+_s\gamma$ and/or $D_s^{*+}\to D^+_s\pi^{0}$ have been theoretically predicted based on effective models,
	e.g.~chiral perturbation theory ($\chi$PT)~\cite{Yang:2019cat,Cheng:1993kp,Cho:1994zu,Wang:2019mhm}, the light-front quark model (LFQM)~\cite{Choi:2007us}, the relativistic quark model (RQM)\cite{Goity:2000dk}, QCD sum rules (QCDSR)~\cite{Aliev:1994nq,Yu:2015xwa},
	the Nambu-Jona-Lasinio model (NJLM)~\cite{Deng:2013uca}, lattice QCD (LQCD)~\cite{Donald:2013sra}, the non-relativistic quark model (NRQM)~\cite{Kamal:1992uv,Fayyazuddin:1993eb},
	and the covariant model (CM)~\cite{Cheung:2015rya}.
	The BF of $D_s^{*+}\to D^+_s \pi^0$ relative to that of $D_s^{*+}\to D^+_s\gamma $ has been measured  by
	using $e^+e^-$ collision data accumulated at the $\Upsilon(3S)$ and $\Upsilon(4S)$ by the CLEO~\cite{cleo2} and BaBar~\cite{babar2} experiments.
	The precision of the world average of the BF of $D_s^{*+}\to D^+_s\gamma$ is about 0.7\%~\cite{pdg2022}.
	Precision measurements of these BFs help to constrain the model parameters, thereby improving the
	effective models. In addition, the BFs are important inputs in the precise
	determination of the $D^+_s$ decay constant $f_{D^+_s}$ and the $c\to s$ CKM matrix element $|V_{cs}|$ via the $e^+e^-\to D_s^{*\pm}D_s^{\mp}$ processes.
	
	In this paper, we report an improved measurement of the BF of $D_s^{*+}\to D^+_s \pi^0$ relative to $D_s^{*+}\to D^+_s \gamma$
	and then determine the BFs of $D_s^{*+}\to  D^+_s\gamma$ and $D_s^{*+}\to D^+_s \pi^0$.
	This analysis is carried out by using 7.33~fb$^{-1}$ of $e^+e^-$ collision data taken at center-of-mass energies $E_{\rm cm}$ between 4.128 and 4.226~GeV with the BESIII detector.

	\section{BESIII detector and Monte Carlo}
	
	The BESIII detector~\cite{Ablikim:2009aa} records symmetric $e^+e^-$ collisions
	provided by the BEPCII storage ring~\cite{Yu:IPAC2016-TUYA01} in the center-of-mass energy range from 2.0 to 4.95~GeV, with a peak luminosity of $1\times10^{33}$~cm$^{-2}$s$^{-1}$ achieved at $\sqrt{s}=3.773~\text{GeV}$.
	BESIII has collected large data samples in this energy region~\cite{Ablikim:2019hff}. BESIII is a cylindrical spectrometer with a geometrical acceptance of $93\%$ over the $4\pi$ solid angle. It consists of a helium-based
	multilayer drift chamber~(MDC), a plastic scintillator time-of-flight
	system~(TOF), and a CsI(Tl) electromagnetic calorimeter~(EMC),
	which are all enclosed in a superconducting solenoidal magnet
	providing a 1.0~T magnetic field. The solenoid is supported by an
	octagonal flux-return yoke with resistive plate counter muon
	identifier modules interleaved with steel~\cite{Huang:geometry}.
	The charged particle momentum resolution is $0.5\%$ at 1 GeV/$c$, and the specific energy loss (d$E$/d$x$) resolution is $6\%$ for the electrons
	from Bhabha scattering. The EMC measures photon energies with a
	resolution of $2.5\%$ ($5\%$) at $1$~GeV in the barrel (end-cap)
	region. The time resolution in the TOF barrel region is 68~ps. The end-cap TOF
	system was upgraded in 2015 using multi-gap resistive plate chamber
	technology, providing a time resolution of
	60~ps~\cite{etof}.  Approximately 83\% of the data used here was collected after this upgrade; luminosities~\cite{lumi} at each energy are given in Table \ref{tab:mbc}.
	
	Simulated data samples are produced with a {\sc
		geant4}-based~\cite{geant4} Monte Carlo (MC) toolkit
	including the geometric description of the BESIII detector and the
	detector response. The simulation includes the beam
	energy spread and initial state radiation (ISR) in the $e^+e^-$
	annihilations with the generator {\sc
		kkmc}~\cite{ref:kkmc}.
	In the MC simulation, the production of open-charm
	processes directly produced via $e^+e^-$ annihilations are modelled with the generator {\sc conexc}~\cite{ref:conexc}.
	The ISR production of vector charmonium(-like) states
	and the continuum processes are incorporated in {\sc
		kkmc}~\cite{ref:kkmc}.
	All particle decays are modelled with {\sc evtgen}~\cite{ref:evtgen} using BFs
	either taken from the
	Particle Data Group~\cite{pdg2022}, when available,
	or otherwise estimated with {\sc lundcharm}~\cite{ref:lundcharm}.
	Final state radiation~(FSR)
	from charged final state particles is incorporated using {\sc photos}~\cite{photos}.
	
	The input cross section line shape of
	$e^+e^-\to D^{*\pm}_sD^{\mp}_s$ is based on the results in Ref.~\cite{crsDsDss}.
	In this analysis, the inclusive MC sample, which is generated at various energy points and has an integrated luminosity of 40 times individual data sets, is used to determine detection efficiencies
	and to estimate background contributions.
	
	\section{Event selection}
	
	At the center-of-mass energies between 4.128 and 4.226~GeV,
	$D^{*+}_s D^-_s$ pairs are produced copiously by $e^+e^-$ collisions.
	The $D^{*+}_s$ mesons decay predominantly via $D^{*+}_s \to D^+_s\gamma$ and $D^{*+}_s \to D^+_s\pi^0$.
	Candidate events are selected by reconstructing $D^+_s$ and $D^-_s$ mesons via hadronic decay modes.
	To obtain better momentum resolution and lower background contamination, we use three modes of
	$D^+_s \to K^+K^-\pi^+$ versus $D^-_s \to K^+K^-\pi^-$,
	$D^+_s \to K^+K^-\pi^+$ versus $D^-_s \to K^0_S K^-$,
	and $D^+_s \to K^0_S K^+$ versus $D^-_s \to K^0_S K^-$,
	which are labelled as modes I, II, and III, respectively.
	In order to improve detection efficiencies, no transition photon or $\pi^0$ from the $D_s^{*+}$ decay is required.
	
	Charged tracks detected in the MDC are required to be within a polar angle ($\theta$) range of $|\rm{cos\theta}|<0.93$, where $\theta$ is defined with respect to the $z$-axis, which is the symmetry axis of the MDC. For charged tracks not originating from $K_S^0$ decays, the distance of closest approach to the interaction point (IP) must be less than 10\,cm along the $z$-axis, $|V_{z}|$, and less than 1\,cm in the transverse plane, $|V_{xy}|$. No additional charged track passing the $\rm{cos\theta}$ and IP cuts is allowed for selected candidates. Particle identification~(PID) for charged tracks combines measurements of the d$E$/d$x$ in the MDC and the flight time in the TOF to form likelihoods for charged pion and kaon hypotheses, $\mathcal{L}(\pi)$ and $\mathcal{L}(K)$.
	Pion candidates are required to satisfy $\mathcal{L}(\pi)$ $>$ $\mathcal{L}(K)$ and $\mathcal{L}(\pi)>0$,
	and kaon candidates are required to satisfy $\mathcal{L}(K)$ $>$ $\mathcal{L}(\pi)$ and $\mathcal{L}(K)>0$.

	The $K_S^0$ candidates are reconstructed via the decay $K^0_S\to \pi^+\pi^-$.
	The two charged pions are required to satisfy $|V_{z}|<20$ cm and $|\!\cos\theta|<0.93$ but no particle identification is applied.
	The $\pi^+\pi^-$ invariant mass is required to be within the interval $(0.487, 0.511)$ GeV/$c^2$.
A vertex fit is performed, constraining the two tracks to originate from a common vertex, and the decay length of $K_S^0$ candidates is required to be greater than twice the resolution.
	
\begin{figure}[!htbp]
		\centering
			\includegraphics[width=0.475\textwidth]{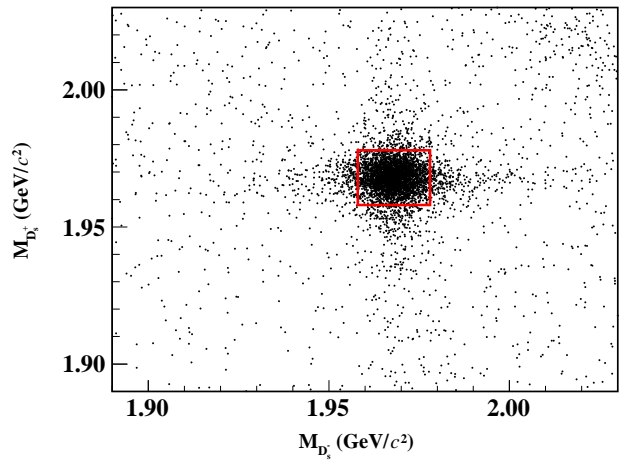}
		\caption{The distribution of $M_{D^-_{s}}$ vs. $M_{D^+_{s}}$ summing over modes I, II, and III in data.
       The red rectangle denotes the signal region.
		\label{com1}}
	\end{figure}

	To suppress non-$D_s^{\pm} D_s^{*\mp}$ events, the beam-constrained mass of the $D_s^-$ candidate
	\begin{equation}
		M_{\rm BC} \equiv  \sqrt {E_{\rm beam}^2-|\vec {p}_{\rm tag}|^2},
	\end{equation}
	is required to be within the intervals as shown in Table~\ref{tab:mbc}. Here, $E_{\rm beam}$ is the beam energy and $\vec p_{\rm tag}$ is the three-momentum of the reconstructed $D_s^-$ candidate in the $e^+e^-$ center-of-mass frame. In each event, we only keep one candidate per tag mode per charge, selecting the one  with the $D_s^-$ recoil mass
	\begin{equation}
		M_{\rm rec} \equiv \sqrt{(E_{\rm cm} - \sqrt{|\vec p_{\rm tag}|^2+m^2_{D_{s}}})^2-|\vec p_{\rm tag}|^2},
	\end{equation}
	closest to the nominal $D_s^{*+}$ mass~\cite{pdg2022}.
		The $D^+_s$ candidate is selected in the presence of the tag $D^-_s$.
		If there are multiple $D^+_s$ combinations in an event, the one giving the minimum $|M_{D_s^+}+M_{D_s^-}-2m_{D_s}|$ is retained for further analysis.
		Here $M_{D^\pm_s}$ is the invariant mass of the $D^\pm_s$ candidate and $m_{D_s}$ is the nominal $D_s$ mass~\cite{pdg2022}.
		Figure~\ref{com1} shows the distribution of $M_{D^-_{s}}$ vs. $M_{D^+_{s}}$ of the accepted candidates in data.
	To suppress background, the invariant masses of $K^+K^-\pi^\pm$ and $K^0_SK^\pm$ combinations
	are required to be within the interval $M_{D^{\pm}_s}\in (1.958, 1.978)$ GeV/$c^2$.
	
	\begin{table}[]
		\caption{The integrated luminosity and $M_{\rm BC}$ requirement for each energy~($E_{\rm cm}$) point.}\vspace{0.2cm}
		\begin{tabular}{ccc}
			\hline
			\hline
			$E_{\rm cm}$ (GeV) & Luminosity (pb$^{-1}$) & $M_{\rm BC}$ (GeV/$c^2$) \\\hline
			4.128             &  401.5 & {[}2.010, 2.061{]} \\
			4.157             &  408.7 & {[}2.010, 2.070{]} \\
			4.178             & 3189.0 & {[}2.010, 2.073{]} \\
			4.189             &  569.8 & {[}2.010, 2.076{]} \\
			4.199             &  526.0 & {[}2.010, 2.079{]} \\
			4.209             &  571.7 & {[}2.010, 2.082{]} \\
			4.219             &  568.7 & {[}2.010, 2.085{]} \\
			4.226             & 1091.7 & {[}2.010, 2.088{]} \\
			\hline
			\hline
		\end{tabular}
		\label{tab:mbc}
	\end{table}

		To improve momentum resolution, a two-constraint (2C) kinematic fit, in which the invariant mass of the $K^+K^-\pi^\pm$ or $K^0_SK^\pm$ combination is constrained to the known $D_s$ mass~\cite{pdg2022} is performed.
		The momenta updated by the kinematic fit are kept for further analysis.
	
\begin{figure}[!htbp]
		\centering
			\includegraphics[width=0.475\textwidth]{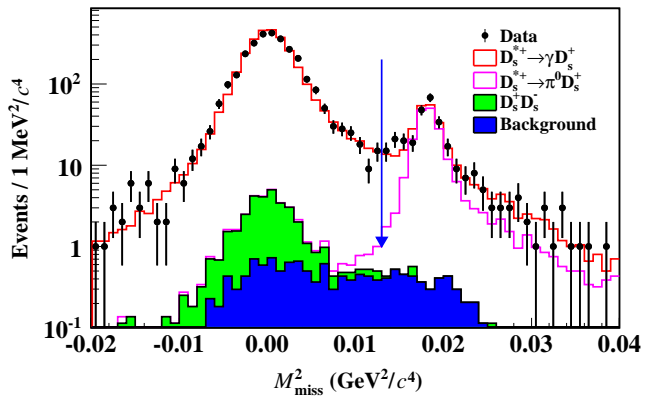}

		\caption{The $M^2_{\rm miss}$ distributions
			of the accepted candidates, summing over
			modes I, II, and III. The points with error bars are data, the
			open red histograms are the scaled signal MC events, and the filled green histograms are normalized
			background events from the inclusive MC sample.
			The blue vertical arrow shows the dividing line for $D^{*+}_s \to D^+_s\gamma$ and $D^{*+}_s \to D^+_s\pi^0$ candidates; for the filled green histogram, the small peaking background around zero is from the $e^+e^-\to D^+_sD^-_s$ process, and the open red and magenta histograms are the signals of $D^{*+}_s \to D^+_s\gamma$ and $D^{*+}_s \to D^+_s\pi^0$, respectively. 		\label{com2}}
	\end{figure}

	To separate the $D^{*+}_s \to D^+_s\gamma$ and $D^{*+}_s \to D^+_s\pi^0$ candidates,
	we define the missing mass squared of the reconstructed $D^+_s D^-_s$ combination as
	\begin{equation}\label{eq3}
		M^2_{\rm miss} \equiv \left (E_{\rm cm}-E_{D^+_s}-E_{D^-_s} \right )^2-|-\vec{p}_{D^+_s}-\vec{p}_{D^-_s}|^2,
	\end{equation}
	where $E_{D^\pm_s}$ and $\vec{p}_{D^\pm_s}$ are the energy and momentum of $D^\pm_s$ in the $e^+e^-$ center-of-mass system, respectively.
	The resultant $M^2_{\rm miss}$ distribution of the accepted $D^+_s D^-_s$ candidate combinations is shown in Fig.~\ref{com2},
	where the peak near to zero and its right-side peak correspond to $D^{*+}_s \to D^+_s\gamma$ and $D^{*+}_s \to D^+_s\pi^0$ candidates, respectively.

	\section{Branching fractions}
	
	Following Ref.~\cite{songwm}, the BF of $D^{*+}_s\to D^+_s \gamma$ relative to the sum of $D^{*+}_s\to D^+_s \gamma$ and $D^{*+}_s\to D^+_s \pi^0$ is determined by
	\begin{equation}
		f_{\gamma} =
		\frac{{\cal B}_{D^{*+}_s\to D^+_s \gamma}}{{\cal B}_{D^{*+}_s\to D^+_s \gamma}+{\cal B}_{D^{*+}_s\to D^+_s \pi^0}} =
		\frac{N_{\gamma}^{\rm prod}}{N_{\gamma}^{\rm prod}+N_{\pi^0}^{\rm prod}},
		\label{eq:bf}
	\end{equation}
	where
	$N_{\gamma}^{\rm prod}$ and $N_{\pi^0}^{\rm prod}$ are the numbers
	of produced $D^{*+}_s\to D^+_s\gamma$ and $D^{*+}_s\to D^+_s\pi^0$
	events, respectively.
This ratio captures the binomial nature of the separation of the
low-background signal into the two decays under study.
	
	As shown in Fig.~\ref{com2}, the individual signal regions of $M_{\rm miss}^{2}$ are defined as $[-0.020, 0.013]$ and $[0.013, 0.040]$ GeV$^2$/$c^4$
	for $D^{*+}_s \to D^+_s\gamma$ and $D^{*+}_s \to D^+_s\pi^0$, respectively.
	The dividing line accepts about 99.0\% of the $D^{*+}\to D_s^+\gamma$ signal and about 98.5\% of the $D_{s}^{*+}\to D_s^+\pi^0$ signal.
	Due to the overlapping $M_{\rm miss}^{2}$ distributions, some $D_{s}^{*+}\to D_s^+\gamma$ events can be misidentified as $D_{s}^{*+}\to D_s^+\pi^0$, and vice versa.
	To account for this effect, the yields of $N_{\gamma}^{\rm prod}$ and $N_{\pi^0}^{\rm prod}$ are obtained by solving the following
	equation
	\begin{equation}
		\left( \begin{array}{l}
			N_{\gamma}^{\rm obs}-N_{\gamma}^{\rm bkg} \\
			N_{\pi^0}^{\rm obs}-N_{\pi^0}^{\rm bkg} \end{array}
		\right) =
		\left( \begin{array}{cc}
			\epsilon_{\gamma\gamma}  & \epsilon_{\pi^0\gamma}  \\
			\epsilon_{\gamma\pi^0} & \epsilon_{\pi^0\pi^0} \end{array}
		\right)
		\left( \begin{array}{l}
			N_{\gamma}^{\rm prod}  \\
			N_{\pi^0}^{\rm prod} \end{array}
		\right),
		\end{equation}
	where $N_{i}^{\rm obs}$ is the number of selected events in data by counting,
	$N_{i}^{\rm bkg}$ is the number of background events estimated from the inclusive MC sample;
	$\epsilon_{ij}$ is the efficiency of the generated
	$D^{*+}_s\to D^+_s+i$ events selected as $D^{*+}_s \to  D^+_s+j$,
	where $i$ and $j$ denote $\gamma$ or $\pi^0$.
        Both $D^{*+}_s \to D_s^+ \pi^0, D_s^+ \gamma$ are simulated.
	The background rates estimated from the inclusive MC sample for modes I, II, and III are all less than 1.5\%.

	To consider different detection efficiencies for ISR and FSR effects,
	the detection efficiencies  at various energy points have been weighted by individual single tag $D^+_s$ yields in data.
	
	Table \ref{tab:BF} lists the quantities used for the $f_{\gamma}$ measurements and the results obtained.
	Weighting the $f_{\gamma}$ results for modes I, II, and III by their inverse statistical uncertainties squared, we obtain their average $f_\gamma = (94.20\pm0.38)\%$.

	\begin{table*}[htbp]\small
		\begin{center}
			\caption{
				The quantities used for $f_\gamma$ measurements and the obtained results.
				The average result is weighted over modes I, II, and III by their inverse statistical uncertainties squared.
				The uncertainties are statistical only.
			}
			\begin{tabular}{ccccccccccc}
				\hline
				\hline
				Mode & $N_{\gamma}^{\rm obs}$ & $N_{\pi^{0}}^{\rm obs}$ & $N_{\gamma}^{\rm bkg}$ & $N_{\pi^{0}}^{\rm bkg}$ &$ \epsilon_{\gamma \gamma} \left(\%\right) $ & $  \epsilon_{\gamma\pi^{0}}\left(\%\right) $ & $ \epsilon_{\pi^{0} \gamma} \left(\%\right) $ & $ \epsilon_{\pi^{0} \pi^{0}} \left(\%\right)$&$ {f}_{\gamma} \left(\%\right) $ \\
				\hline
				I & $2293.0\pm47.9$&$239.0\pm15.5$&$31.0\pm0.9$&$5.0\pm0.4$   &$14.16\pm0.04$ &$0.42\pm0.01$ &$0.22\pm0.02$ &$15.08\pm0.17$ &$93.52\pm0.49$ \\
				II& $1044.0\pm32.3 $&$83.0\pm9.1 $&$12.0\pm0.5$&$1.0\pm0.2$   &$15.97\pm0.07$ &$0.46\pm0.01$ &$0.16\pm0.03$ &$16.38\pm0.29$ &$95.32\pm0.63$ \\
				III& $119.0\pm10.9 $&$11.0\pm3.3  $&$1.0\pm0.2$&$0.0\pm0.0$   &$17.27\pm0.23$ &$0.52\pm0.05$ &$0.00\pm0.00$ &$18.08\pm0.96$ &$94.31\pm2.04$ \\
				\hline
				Average  &       &     &     &     &  & &          &                &$94.20\pm0.38$ \\
				\hline
				\hline
			\end{tabular}
			
		\label{tab:BF}
	\end{center}
\end{table*}

\section{Systematic uncertainties}

The systematic uncertainties in the BF measurements are discussed below.
The systematic uncertainty due to $M_{{\rm miss}}^2$ resolution is examined in the following procedure.
We perform a fit to $M_{{\rm miss}}^2$ distribution of data. To take into account the resolution difference between data and MC, a signal MC shape smeared with a Gaussian function is used.
From the fit, we obtain the parameters~(means, widths) of the Gaussian resolution functions, which are
$(1.0\pm0.1,1.0\pm0.2)$, $(1.1\pm0.1,1.0\pm0.2)$, and $(0.4\pm0.3,1.7\pm0.4)$ MeV$^2$/$c^4$ for modes I, II, and III, respectively.
The change of BF before and after smearing the Gaussian resolution function to the $M_{\rm miss}^{2}$ distribution of the signal MC events, 0.07\%, is taken as the associated systematic uncertainty.

The systematic uncertainty caused by the statistical uncertainty of the MC efficiencies is estimated by varying
each of the efficiency matrix elements by $\pm 1\sigma$.
The largest change of the BF is taken as the systematic uncertainty.

The systematic uncertainty from background estimation is considered in two parts.
The number of background events is calculated from the inclusive MC
sample. The corresponding systematic uncertainty is estimated from the uncertainties of the cross sections used in generating this sample. The dominant background events are from open charm processes of $e^+e^-\to D^+_sD^-_s$ and $e^+e^-\to D^{*+}_sD^-_s$. The systematic uncertainty is estimated by varying
the cross sections and BFs of the hadronic $D^+_s$ decays by $\pm 1\sigma$.
This effect on the BF measurement is negligible.
In addition, we have also varied the simulated background events by the ratio of the background events
observed in the $D^+_sD^-_s$ sideband regions between data and the inclusive MC sample.
The change of the BF, 0.10\%, is taken as the corresponding systematic uncertainty.

Other possible systematic uncertainty sources, such as the ISR simulation,
the kinematic fit, the tracking and the particle identification efficiencies between the two decay modes of $D_s^{*+}$, the $M_{\rm BC}$ requirement and the $M^2_{\rm miss}$ range, have also been investigated.
All of them are negligible.

All systematic uncertainties are summarized in Table~\ref{tab:summary}.
Assuming the systematic uncertainties from different sources are independent,
the total systematic uncertainty is obtained to be 0.17\% by adding all the sources quadratically.

\section{Summary}

By analyzing 7.33 fb$^{-1}$ of $e^+e^-$ collision data taken at center-of-mass energies between
4.128 and 4.226~GeV,
we measure the BF of $D^{*+}_s\to D^+_s \gamma$ relative to the sum of $D^{*+}_s\to D^+_s \gamma$ and $D^{*+}_s\to D^+_s \pi^0$ to be $f_\gamma=(94.20\pm 0.38\pm 0.17)\%$. This gives the BF of $D^{*+}_s\to D^+_s\pi^0$ relative to that of $D^{*+}_s\to D^+_s\gamma$ to be
${\cal B}_{D^{*+}_s \to D^+_s\pi^0}/{\cal B}_{D^{*+}_s \to D^+_s\gamma} = \frac{1}{f_\gamma}-1= (6.16\pm 0.43\pm 0.19)\%$.
The $D^{*+}_s$  is known to decay dominantly into three final states of $D^+_s\gamma$, $D^+_s\pi^0$ and $D^+_se^+e^-$~\cite{Cheung:2015rya}. Combining
the world average of  ${\cal B}_{D^{*+}_s\to D^+_se^+e^-}=(0.67\pm0.16)\%$~\cite{pdg2022}, we obtain
${\cal B}_{D^{*+}_s\to D^+_s \gamma}=(93.57\pm 0.44\pm 0.19)\%$ and ${\cal B}_{D^{*+}_s\to D^+_s \pi^0}=(5.76\pm 0.44\pm 0.19)\%$.

\begin{figure}[htbp]
	\centering
	\includegraphics[width=0.5\textwidth]{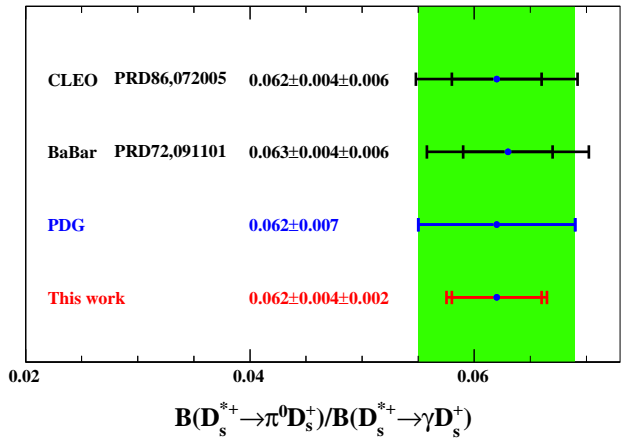}
	\caption{Comparison of ${\cal B}_{D^{*+}_s \to D^+_s\pi^0}/{\cal B}_{D^{*+}_s \to D^+_s\gamma}$ measured by this work and previous experiments.
		The points with error bars are from different experiments.
		For each experiment, the shorter error bar denotes statistical only
		while the longer error bar combines both statistical and systematic uncertainties.
		The green band corresponds to the $\pm 1\sigma$ limit of the world average.
	} \label{com_all}
\end{figure}

Figure~\ref{com_all} shows the comparison of the measured BF of ${\cal B}_{D^{*+}_s \to D^+_s\pi^0}/{\cal B}_{D^{*+}_s \to D^+_s\gamma}$ with other experiments and the
world average value~\cite{pdg2022}. Our measurement is well consistent with the
previous ones but with better precision.
Table \ref{tab:com} shows comparisons of the BFs measured in this work with the world average values and the decay widths or BFs predicted by various theories.
Our results of ${\mathcal B}_{D^*_s \to D_s^+\gamma}$ and ${\mathcal B}_{D^*_s \to D_s^+\pi^0}$ are consistent with those predicted in Ref.~\cite{Cheung:2015rya}.
At present, only limits on the $D^{*+}_s$ width have been reported.
More experimental measurements and theoretical calculations of the $D^{*+}_s$ decays will be beneficial to
give quantitative tests on the predicted partial decay widths, thereby better understand the radiative and strong decays of $D^{*+}_s$. As necessary inputs, the reported BFs with much improved precision are also important for the precise measurements of $f_{D^+_s}$ and $|V_{cs}|$ by using the reactions of $e^+e^-\to D_s^{*\pm}D_s^{\mp}$.

\begin{table}[htbp]
	\begin{center}
		\caption{Relative systematic uncertainties in the determination of $f_\gamma$.}\vspace{0.2cm}
		\begin{tabular}{lc}
			\hline
			\hline
			Source                          &Uncertainty~$\left(\% \right)$ \\ \hline
			$M_{{\rm miss}}^2$ resolution & 0.07 \\
			MC statistics                 & 0.12 \\
			Background                    & 0.10\\
			\hline
			Sum & 0.17\\
			\hline
			\hline
		\end{tabular}
		\label{tab:summary}
	\end{center}
\end{table}

\section*{acknowledgements}
The BESIII collaboration thanks the staff of BEPCII and the IHEP computing center for their strong support. This work is supported in part by National Key R\&D Program of China under Contracts Nos. 2020YFA0406400, 2020YFA0406300; National Natural Science Foundation of China (NSFC) under Contracts Nos. 11635010, 11735014, 11835012, 11935015, 11935016, 11935018, 11961141012, 12022510, 12025502, 12035009, 12035013, 12192260, 12192261, 12192262, 12192263, 12192264, 12192265; the Chinese Academy of Sciences (CAS) Large-Scale Scientific Facility Program; Joint Large-Scale Scientific Facility Funds of the NSFC and CAS under Contract No. U1832207, U1932102; 100 Talents Program of CAS; The Institute of Nuclear and Particle Physics (INPAC) and Shanghai Key Laboratory for Particle Physics and Cosmology; ERC under Contract No. 758462; European Union's Horizon 2020 research and innovation programme under Marie Sklodowska-Curie grant agreement under Contract No. 894790; German Research Foundation DFG under Contracts Nos. 443159800, Collaborative Research Center CRC 1044, GRK 2149; Istituto Nazionale di Fisica Nucleare, Italy; Ministry of Development of Turkey under Contract No. DPT2006K-120470; National Science and Technology fund; National Science Research and Innovation Fund (NSRF) via the Program Management Unit for Human Resources \& Institutional Development, Research and Innovation under Contract No. B16F640076; STFC (United Kingdom); Suranaree University of Technology (SUT), Thailand Science Research and Innovation (TSRI), and National Science Research and Innovation Fund (NSRF) under Contract No. 160355; The Royal Society, UK under Contracts Nos. DH140054, DH160214; The Swedish Research Council; U. S. Department of Energy under Contract No. DE-FG02-05ER41374.

\begin{table*}[htbp]
\begin{center}
\caption{Comparisons of the partial widths ($\Gamma$) and BFs (in brackets).  The decay widths are in units of keV. The first two rows are from this work and the PDG, while the others are from various theoretical predictions.
The superscript $^a$ denotes the value corresponding to $g=0.52$, $\beta=2.6$ GeV$^{-1}$, and $m_c=1.6$ GeV;
$^b$ denotes the values for a linear model;
$^c$ denotes the value for $\kappa^q=0.55$; and
$^d$ denotes the values for $(a)$ model.
\label{tab:com}}
\begin{tabular}{cccc} \hline \hline
                 & $ \Gamma\,[\mathcal B]_{D^*_s \to D_s^+\gamma}$  & $ \Gamma\,[\mathcal B]_{D^*_s \to D_s^+\pi^0}$  & ${\cal B}_{D^{*+}_s \to D^+_s\pi^0}/{\cal B}_{D^{*+}_s \to D^+_s\gamma}$   \\ \hline
This work                          &...$[(93.57\pm 0.41\pm 0.16)\%]$  &...$[(5.76\pm 0.39\pm 0.16)\%]$               & $(6.16\pm 0.40\pm 0.17)\%$ \\
PDG~\cite{pdg2022}                 & ...$[(94.2\pm 0.7)\%]$           &...$[(5.9 \pm 0.7)\%]$                        &  $(6.2\pm0.8)\%$ \\ \hline
CM~\cite{Cheung:2015rya}           & 3.53 $[(92.7 \pm 0.7)\%]$        & $0.277^{+0.028}_{-0.026}$ $[(7.3\pm 0.7)\%]$ &   $(7.9\pm0.8)\%$ \\
$\chi$PT \cite{Cheng:1993kp}$^a$   & $4.5$                            & ...                                          & ... \\
$\chi$PT \cite{Cho:1994zu}         & ...                              & ...                                          & $8 \times 10^{-5}/{\cal B}(D^{*+} \to D^{+} \gamma)$ \\
$\chi$PT \cite{Wang:2019mhm}       & $0.32\pm0.30$                    & ...                                          & ... \\
$\chi$PT \cite{Yang:2019cat}       & ...                              & $0.0081^{+0.0030}_{-0.0026}$                 & ... \\
LFQM \cite{Choi:2007us}$^b$        & $0.18\pm 0.01$                   & ...                                          & ... \\
RQM \cite{Goity:2000dk}$^c$        & $0.321^{+0.009}_{-0.008}$        & ...                                          & ... \\
QCDSR \cite{Aliev:1994nq}          & $0.25\pm 0.08$                   & ...                                          & ... \\
QCDSR \cite{Yu:2015xwa}            & $0.59\pm 0.15$                   & ...                                          & ... \\
NJLM \cite{Deng:2013uca}           & $0.09$                           & ...                                          & ... \\
LQCD \cite{Donald:2013sra}         & $0.066\pm 0.026$                 & ...                                          & ... \\
NRQM \cite{Kamal:1992uv}           & $0.21$                           & ...                                          & ... \\
NRQM \cite{Fayyazuddin:1993eb}$^d$ & $0.40$                           & ...                                          & ... \\ \hline \hline
\end{tabular}
\end{center}
\end{table*}

\end{document}